# ИНФОРМАТИЗАЦИЯ ПРОЕКТНЫХ РАБОТ НА ПРОМЫШЛЕННОМ ПРЕДПРИЯТИИ ПРИ ЕГО РЕКОНСТРУКЦИИ

Россия, г.Казань, Центр экономических и социальных исследований Республики Татарстан при Кабинете Министров Республики Татарстан
Мигунов В.В., к.т.н., нач.отдела

## Введение

Безусловно необходимый элемент успешной деятельности любого производственного предприятия, действующего в условиях рыночной экономики - постоянное развитие, освоение новой продукции, технологий и оборудования, что включает непрерывное решение задач реконструкция предприятия. Статья посвящена процессам информатизации соответствующего вида профессиональной деятельности - проектирования реконструкции действующего промышленного предприятия. Эти работы следует отличать от сильно специализированных задач проектирования самой продукции предприятий, например, вертолетов или бытовых приборов. При реконструкции объектами проектирования являются здания и сооружения, технологические установки, системы водо-, тепло-, электроснабжения и т.п., которые в различных отраслях промышленности имеют много общего. Практически на всех средних и крупных, и даже на многих малых предприятиях имеется необходимость в выполнении проектов реконструкции предприятия собственными силами. Проекты выполняются проектно-конструкторскими подразделениями с различным статусом и объемом работ - от одного конструктора до проектного института. Типовой, усредненной организационной формой можно считать проектно-конструкторский отдел (ПКО) предприятия. Содержание профессиональной деятельности конструкторов, технологов, инженеров ПКО имеет ряд особенностей, отражающихся, в частности, и в технологиях информатизации этой деятельности.

В настоящей работе дается характеристика проектных работ на предприятии при его реконструкции, общих направлений их информатизации, выделяется ключевое направление - внедрение систем автоматизированного проектирования (САПР), формулируются требования к комплексной САПР, описываются методы и средства развития такой САПР, характеризуется связь этого развития с интеграцией информационного пространства ПКО. В основе рассмотрения лежит десятилетний опыт разработки и внедрения отечественной комплексной САПР реконструкции предприятий TechnoCAD GlassX [1].

## 1. Характеристика проектных работ в ПКО предприятия

Основанием для выполнения проектов в ПКО являются задания на проектирование, поступающие в связи с планами реконструкции, изменением состава и объема выпускаемой продукции, предписаниями Госгортехнадзора, возникновением аварийных ситуаций и др.

Результатом проектирования являются графические (в основном) и текстовые конструкторские документы для строительно-монтажных работ, объединяемые понятием проектно-сметной документации. Нет возможностей сквозного использования электронного проекта вплоть до изготовления изделий, какие возникают, например, в машиностроении и электронике.

Основную трудоемкость составляет разработка чертежей. Все графические документы подчиняются требованиям одновременно двух систем российских стандартов - ЕСКД и СПДС (единая система конструкторской документации и система проектной документации для строительства), что часто затрудняет применение иностранных графических пакетов.

Реконструкцию характеризует небольшой объем каждого проекта при большом разнообразии марок входящих в него чертежей. Варианты марок ТХ, ТК, ГСН, ГТ, ГП, АР, КЖ, КМ, КД, ОВ, ВК, НВК, ТС, ЭМ, ЭО, ЭН, ЭС, А.. встречаются в одном проекте в количестве нескольких штук.

В чертежах различных марок имеются общие части, такие как строительная подоснова или технологическая схема. Реально производится последовательная совместная итеративная разработка чертежей, в том числе в процессах согласования с пожарной инспекцией и другими службами.

Подавляющая часть проектной документации на предприятия, построенные до недавнего времени, имеется в основном в бумажной форме. Перевод ее в векторный графический формат имеет смысл проводить лишь постепенно, по мере необходимости, которая и возникает в проектах реконструкции.

Выполнение большого количества (несколько десятков или сотен) проектов в течение года силами ПКО с небольшой численностью сотрудников затрудняет специализацию рабочих мест. На каждом рабочем месте разрабатываются чертежи нескольких марок и требуются соответствующие САПР.

По своей природе деятельность проектировщика совмещает в себе использование собственных общих технических и технологических знаний, знаний о собственном предприятии, стандартов ЕСКД и СПДС, каталогов выпускаемых промышленностью изделий, программного обеспечения для расчетов, разработки чертежей, для поиска в базах данных. Он работает с информацией, представленной на бумаге или в электронном виде. Как и в других видах профессиональной деятельности, связанных исключительно с обработкой информации, главным способом повышения производительности труда проектировщика является информатизация "в узком смысле" [2], то есть внедрение информационных технологий, упрощающих и ускоряющих доступ к информации.

Для целей настоящего рассмотрения условно разделим процессы информатизации проектных работ на общие и специфичные для этой профессиональной деятельности.

## 2. Общая информатизация деятельности проектировщика

К процессам общей информатизации отнесем проникновение в деятельность проектировщика информационных технологий, реализующих, в частности:

• быстрый доступ к информации на рабочем месте с помощью компьютера;

• ускорение и прототипирование подготовки текстовых документов в системе делопроизводства, включая расчетно-пояснительные записки, с помощью редактора текстов (обычно MS Word);

- ускорение и прототипирование расчетов, графическое оформление результатов с помощью электронных таблиц (обычно MS Excel);
- обмен данными, документами и чертежами с другими рабочими местами и подразделениями посредством локальных компьютерных сетей;
- использование электронных сборников ГОСТов, ОСТов, СНиПов и других нормативных документов с соответствующим повышением степени их актуальности;
- обмен данными и чертежами с другими подразделениями предприятия и с другими предприятиями по электронной почте;
- применение Internet для поиска поставщиков оборудования и услуг, их каталогов и price-листов;
- повышение качества документов и чертежей в бумажном виде за счет все возрастающих возможностей печатающих устройств.

Этот список не претендует на полноту, но дает представление о процессах информатизации, которые не специфичны для ПКО, а могут быть отнесены и к другим подразделениям предприятия. На фоне этих процессов идет информатизация и сугубо специфичной деятельности ПКО.

### 3. Информатизация специфичной деятельности проектировщика

Основное содержание этой деятельности - проектирование. С этих позиций и проводится выделение специфичных процессов информатизации:
- исключаются относительно общие процессы информатизации планирования проектных работ, учета и контроля, подготовки сметной документации;
- исключаются информационные технологии, отнесенные выше к общей информатизации;
- под информатизацией понимается прежде всего внедрение современных информационных технологий, выражающееся в последовательном наращивании степени автоматизации проектирования;
- среди программного обеспечения, отвечающего этой задаче, выделяется то, которое связано с обработкой общих частей проектов на нескольких рабочих местах, то есть с наибольшими потоками обмена информацией. При этом из поля зрения выпадает большое число специализированных расчетных программ, таких, как расчет освещенности, прочностной и гидравлический расчет трубопроводов, расчет предохранительных клапанов и др.

В результате специфичная информатизация деятельности проектировщика выражается во внедрении САПР, используемой на нескольких рабочих местах и решающей задачи, вытекающие из вышеприведенной характеристики проектных работ в ПКО предприятия. Далее следует краткий анализ этих задач и соответствующие требования к такой САПР.

### 4. Требования к комплексной САПР реконструкции

Необходимы САПР с единым пользовательским интерфейсом, иначе велики потери времени на освоение разнообразных приемов работы в нескольких средах проектирования.

Большинство проектов выполняются "поверх" прежних чертежей, с изменением одной части и сохранением другой. Поскольку прежние чертежи хранятся в архивах ПКО в бумажном виде, варианты приемлемых графических пакетов ограничиваются теми, которые обеспечивают эффективную работу с отсканированными чертежами одновременно с векторной частью (гибридные редакторы).

Чертежи должны передаваться от одного проектировщика к другому и допускать доработку на разных рабочих местах, что возможно при поддержке единого графического формата.

Процесс автоматизации проектных работ в ПКО действующего предприятия идет постепенно. В том числе и в части технического оснащения. Сначала – постепенное оснащение компьютерами рабочих мест, где компьютеров не было. Затем – постоянное переоснащение новыми с параллельным оснащением изменяющимися моделями устройств печати. На фоне этой постепенности желательно, чтобы САПР работали на компьютерах минимальной из имеющихся конфигураций и с самыми разнообразными устройствами печати.

Таким образом, задача автоматизации проектирования реконструкции предприятия требует программного обеспечения, ориентированного на выпуск графической части проектов по требованиям ЕСКД и СПДС, работающего одновременно с растровой и векторной графикой, в едином пользовательском интерфейсе автоматизирующего подготовку чертежей различной проблемной ориентации в едином графическом формате, с минимальной требовательностью к ресурсам.

Кроме этих, достаточно общих и относящихся непосредственно к САПР требований, для ее внедрения требуются еще некоторые условия, относящиеся к взаимосвязанным процессам ее внедрения и развития.

## 5. Технология внедрения и развития комплексной САПР реконструкции

Чтобы автоматизация проектных работ приносила реальную пользу, требуется заинтересовать главное действующее лицо этой автоматизации – проектировщика, то есть создать условия, в которых работники ПКО сами будут стремиться применять САПР.

Основной принцип решения задачи автоматизации работ в этих условиях состоит не в разовом установлении требований к программному обеспечению и технике, а в установлении требований к самому процессу постепенного перехода к компьютерным информационным технологиям в проектировании конкретно в ПКО. И лишь во вторую очередь из общей идеологии процесса перехода вытекают частные требования к программному обеспечению, причем они изменяются в зависимости от стадии процесса.

В результате десятилетних усилий выработаны приемлемые подходы, которые условно можно разделить на 4 направления: самораспространение САПР, самостоятельное освоение, оперативное сопровождение разработчиком и непрерывное развитие САПР. Ниже они описываются подробнее вместе с особенностями САПР, реализованными в TechnoCAD GlassX.

## 5.1. Самораспространение

Отношения с разработчиком (поставщиком) строятся таким образом, что у ПКО и у предприятия в целом нет ограничений на число рабочих мест, где применяется САПР. Программы нетребовательны к ресурсам и работают на любом компьютере, который встречается в ПКО, включая различные операционные системы. Отсутствует защита от копирования. Любой проектировщик (конструктор) может самостоятельно установить TechnoCAD GlassX и поработать с ним, оценить удобства и качество результатов.

## 5.2. Самостоятельное освоение

Изначально русскоязычный пользовательский интерфейс, полная поддержка требований привычных отечественных стандартов ЕСКД и СПДС, отсутствие ненужных (и поэтому непонятных и мешающих) возможностей, контекстная справка в любом состоянии ожидания – вот основные свойства TechnoCAD GlassX, приводящие к тому, что все пользователи освоили его практически самостоятельно, изредка задавая вопросы по телефону. Несколько труднее он осваивался работниками, которые не занимаются проектированием на постоянной основе, но вносят изменения в технологические схемы регламентов производства. Здесь помог вводный курс обучения.

Отсутствие ненужных возможностей при подготовке чертежей определенных марок достигается заданием так называемого "профиля работ" (рис.1). Каждому профилю работ соответствует свой набор опций главного меню.

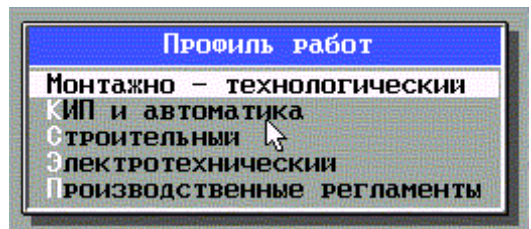

Рис.1. Выбор профиля проектных работ

## 5.3. Оперативное сопровождение автоматизации работ в ПКО

Оперативное сопровождение включает, прежде всего, консультации по телефону в рабочее время, передачу новых версий на конкретные рабочие места по электронной почте. Также выполняется ряд работ непосредственно в ПКО или на предприятии: подключение и пусконаладка новых устройств печати или сканирования, массовая установка новых версий и демонстрация их возможностей, разрешение вновь возникающих и ранее встречавшихся, но забытых программных и технических конфликтов, обсуждение планов технического оснащения ПКО, развития сетей.

## 5.4. Непрерывное развитие САПР

Развитие САПР TechnoCAD GlassX осуществляется двумя путями.

Первый можно назвать эволюционным. Осуществляется адаптация к новым инженерным условиям эксплуатации. В частности, к вводу в действие но-вых ГОСТов, ОСТов, стандартов предприятия, руководящих технических материалов, номенклатурных каталогов изделий, к появлению новых операционных систем, устройств печати, к изменению минимальных конфигураций имеющихся компьютеров, к появлению и изменениям сетевых возможностей взаимодействия рабочих мест.

Второй путь – включение новых проблемно-ориентированных расширений, в том числе комплексирование с другими программными системами. По мере освоения и привыкания к уже имеющимся возможностям TechnoCAD GlassX и системе его поддержки и адаптации у пользователей возникают новые идеи по автоматизации специфичных работ, в том числе и не только проектных. После обсуждения с разработчиком определяется постановка задачи и делается заказ на разработку. Примеры: подготовка схем автоматизации технологических процессов; специфицирование аксонометрических схем паропроводов и трубопроводов высокого давления, подведомственных Госгортехнадзору; программный комплекс ведения в электронной форме регламентов производства, состоящих из текстовых частей в формате MS Word и из технологических схем в формате TechnoCAD GlassX. В результате разработки ряда специализированных расширений сложилась эффективная технология их создания.

### 5.5. Технология создания специализированных расширений

Для САПР, направленных на создание чертежей, разработка специализированных проблемно-ориентированных расширений оправдана в случаях, когда за счет их применения трудоемкость ввода данных существенно ниже трудоемкости непосредственного черчения с помощью графического ядра САПР, включая проведение необходимых расчетов. Например, когда требуются трудоемкие расчеты или когда нормативные требования к чертежу порождают большое количество графических изображений по малому количеству исходных данных. В обоих случаях наиболее предпочтительна параметрическая генерация чертежа по исходным данным. Для решения задач такого класса применяется технология, основанная на совместном хранении в одном элементе чертежа, называемом "Модуль", как исходных данных, так и результатов параметрической генерации.

Модуль включает видимую в чертеже совокупность геометрических элементов (рис.2) и невидимое в чертеже параметрическое представление моделируемого объекта. Третья часть проблемно-ориентированного расширения - процедуры работы с модулем, они помещаются в программный код TechnoCAD GlassX.

Параметрические представления объектов в модулях вместе со специализированным кодом расширения САПР позволяют реализовывать самые разные модели объектов и методов их разработки. Специализация модулей легко опознается в чертеже, и возникает возможность многоэтапной разработки объектов

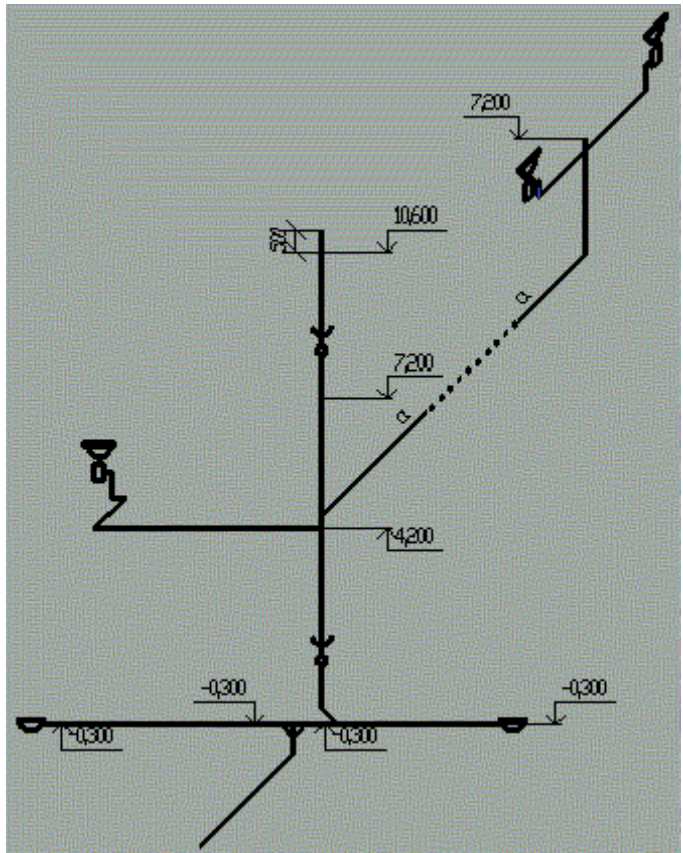

Рис.2. Модуль аксонометрической схемы трубопроводной системы проектирования и использования объектов - прототипов. Высокая структурированность параметров обеспечивает быстрый доступ к ним для различной обработки. Например, в модуль аксонометрической схемы можно автоматически вставить оси строительной подосновы, лишь выбрав модуль подосновы в чертеже. Легко автоматизируется сбор сведений при генерации спецификаций, при контроле дублирования позиционных обозначений.

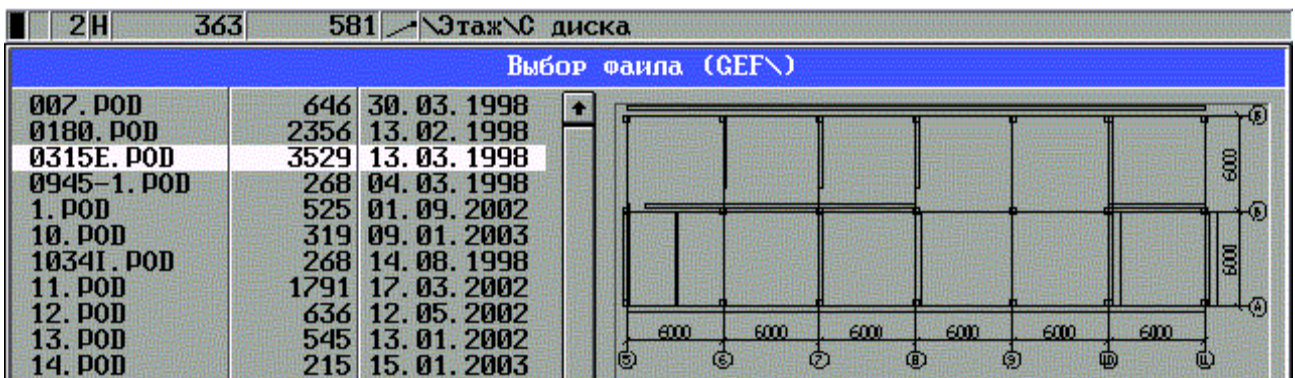

Рис.3. Выбор прототипа из дисковых комплектов параметров строительной подосновы

Комплект параметров модулей может записываться на диск (без геометрических элементов), порождая информационную среду проектирования в виде библиотек прототипов. При выборе комплекта для чтения с диска геометрия генерируется в режиме on-line, и проектировщик легко ориентируется в прототипах (рис.3).

Как элемент чертежа модуль подчиняется обычным правилам: его можно удалить, подвинуть, растянуть и т.д., к его геометрическим элементам возмож-ны привязки, он может быть помещен в графическую библиотеку.

При задании параметрического представления используется как стандартный пользовательский интерфейс (меню, формы ввода...), так и специализированный для черчения и корректировки.

Таким образом, проектировщик по мере развития TechnoCAD GlassX всегда остается в привычном ему пользовательском интерфейсе. Специфику работы с модулями освоили уже все пользователи, и при добавлении проблемных расширений проектировщику остается освоить конкретную реализацию задачи, которую он решает каждодневно и которая для него глубоко понятна.

### 7. Заключение. Интеграция информационного пространства ПКО

Процессы компьютеризации, объединения рабочих мест в локальные и корпоративные сети с выходом в глобальные постепенно создают все более интегрированное электронное информационное пространство. Представляет интерес проследить, какие изменения в ходе этих процессов претерпевает САПР TechnoCAD GlassX.

Начальный этап. Программа используется на нескольких рабочих местах монтажно-технологического бюро проектно-конструкторского отдела (ПКО). Чертежи при необходимости передаются на дискетах.

По мере компьютеризации рабочих мест эта САПР начинает применяться в других подразделениях ПКО: бюро автоматизации, электротехническом, строительном и сантехническом бюро.

Параллельно с появлением локальной сети развиваются: САПР схем автоматизации, в которую передаются технологические схемы из монтажно-технологического бюро; САПР строительной подосновы и обмен ее чертежами между всеми бюро по локальной сети; импорт схем САПР-Альфа, TechnoCAD Elec, TechnoCAD Power для нужд электротехнического бюро; разрабатывается редактор специальных стилей штрихования для чертежей строительного профиля; ввиду многообразия функций вводятся профили работ, сужающие спектр возможностей до нужных каждому бюро в отдельности.

Возникает возможность вести управление проектной документацией и проектными работами в локальной сети - появляется соответствующая программа, в которой часть программного кода TechnoCAD GlassX используется для просмотра чертежей.

В техническом отделе предприятия после появления компьютеров и решения первоочередных задач учета потребовался переход к ведению регламентов производства в электронной форме. TechnoCAD GlassX доработан для автоматизированной подготовки технологических схем производства в соответствующих стандартах, а его часть, обеспечивающая просмотр чертежей, стала частью кода интегратора электронных регламентов (работающего также и с текстовыми частями регламентов) и частью кода программы просмотра электронных регламентов с CD дисков на внешних компьютерах.

Интенсификация обмена чертежами потребовала защиты и контроля их целостности. В TechnoCAD GlassX эта задача решена методом электронной подписи с автообновлением ключевых файлов по локальной сети.

По мере выхода в глобальные сети передача новых версий программного обеспечения осуществляется по электронной почте, и время от обнаружения ошибки в программах до установки версии без этой ошибки составляет иногда несколько часов.

## Список источников